\documentclass[prc,aps,amssymb,twocolumn,showpacs]{revtex4} 

\usepackage{aas_macros} 
\usepackage{hyperref}   
\usepackage{natbib}     
\usepackage{subfigure}  
\usepackage{graphicx}   
\usepackage{longtable}  

\newcommand{\nggn}{$(n,\gamma)\rightleftarrows(\gamma,n)$}  

\begin{document}

\title{The impact of individual nuclear masses on $r$-process abundances}
\author{M.R. Mumpower$^{1,2}$}
\author{R. Surman$^{1,2}$}
\author{D.-L. Fang$^{2,3}$}
\author{M. Beard$^{1,2}$}
\author{P. M{\"o}ller$^{4}$}
\author{T. Kawano$^{4}$}
\author{A. Aprahamian$^{1,2}$}
\affiliation{$^{1}$Department of Physics, University of Notre Dame, Notre Dame, Indiana 46556, USA}
\affiliation{$^{2}$Joint Institute for Nuclear Astrophysics, USA}
\affiliation{$^{3}$Department of Physics, Michigan State University, East Lansing, MI 48824, USA}
\affiliation{$^{4}$Theoretical Division, Los Alamos National Laboratory, Los Alamos, New Mexico 87545, USA}

\date{\today}

\begin{abstract} 
We have performed for the first time a comprehensive study of the sensitivity of $r$-process nucleosynthesis to 
individual nuclear masses across the chart of nuclides. Using the latest version (2012) of the Finite-Range Droplet 
Model, we consider mass variations of $\pm0.5$ MeV and propagate each mass change to all affected quantities, 
including $Q$-values, reaction rates, and branching ratios. We find such mass variations can result in up to an order 
of magnitude local change in the final abundance pattern produced in an $r$-process simulation. We identify key 
nuclei whose masses have a substantial impact on abundance predictions for hot, cold, and neutron star merger 
$r$-process scenarios and could be measured at future radioactive beam facilities.
\end{abstract}


\maketitle

\section{Introduction} 
\label{sec:intro}

One of the most challenging open questions in all of physics is the identification of the site or sites of 
rapid neutron capture, or $r$-process, nucleosynthesis \cite{BBFH1957,NRC2002}. Production of the heaviest 
$r$-process elements requires on the order of 100 neutron captures per seed nucleus; exactly where and how such rapid 
neutron captures occur has yet to be definitively determined \cite{Arnould2007}.

One attractive potential site is within the cold or mildly heated tidal ejecta from neutron star or neutron 
star-black hole mergers \cite{Lattimer+74}. Current state-of-the-art simulations show a vigorous $r$ process with 
fission recycling in the merger ejecta \cite{Goriely2011,Korobkin2012,Wanajo+14,Just+15}. The resulting abundance 
pattern is relatively insensitive to variations in the initial conditions, which naturally explains the consistent 
$56<Z<82$ pattern observed in the solar system and $r$-process-enhanced halo stars \cite{Sneden2008,Roederer+14}. 
Vigorous production of radioactive $r$-process nuclei can also lead to an observable electromagnetic transient 
accompanying the merger event \cite{Li+98,Roberts+11,Barnes+13}, an example of which may have already been detected 
\cite{Berger2013,Tanvir+13}. It is less clear whether mergers happen often enough or early enough in galactic history 
to fit all of the observational data \cite{Argast+04,Komiya+14,Matteucci+15}. The neutrino-driven wind within a 
core-collapse supernova is perhaps the best-studied alternative \cite{Meyer+92,Woosley+94}, though the combination of 
moderate neutron-richness, high entropy, and fast outflow timescale required to make the heaviest $r$-process 
elements does not appear to be achieved in modern simulations 
\cite{Hudepohl2010,Fischer2010,Arcones2013,Roberts2012}. Instead attention has shifted to more exotic sites connected 
to the deaths of massive stars, including neutron-rich jets \cite{Winteler+12}, supernova neutrino-induced 
nucleosynthesis in the helium shell \cite{Banerjee+11}, and collapsar outflows \cite{Nakamura+13,Malkus+12}.

In principle, the proposed environments have such distinct astrophysical conditions that their abundance pattern 
predictions should be clearly distinguishable. Currently simulations lack this precision 
\cite{Mumpower2014_CGS15,Mumpower2014F}, in large part due to uncertainties in the required nuclear data. Properties 
such as masses, neutron capture rates, $\beta$-decay rates, and $\beta$-delayed neutron emission probabilities are 
needed for thousands of neutron-rich nuclear species from the valley of stability to the neutron drip line. Presently 
there is little experimental information available for the vast majority of these quantities. 
Simulations must instead rely on extrapolated or theoretical values, where different approaches can produce markedly different (and often divergent) predictions.

Nuclear masses are particularly important for the $r$ process as they enter into the calculations of all of the 
aforementioned nuclear properties, which shape how each phase of the $r$ process proceeds. In a classic $r$ process, 
an equilibrium is established between neutron captures and photodissociations, and nuclear masses directly determine 
the $r$-process path through a Saha equation:
\begin{equation}
\frac{Y(Z,N+1)}{Y(Z,N)}\propto\frac{G(Z,N+1)}{2G(Z,N)}\frac{N_{n}}{(kT)^{3/2}} \exp\left[\frac{S_{n}(Z,N+1)}{kT}\right]
\label{eq:saha}
\end{equation}
where $G(Z,N)$ are the partition functions, $N_{n}$ is the neutron number density, $kT$ is the temperature in MeV, 
and $S_n(Z,N+1)$ is the neutron separation energy, the difference in binding energy between the nuclei $(Z,N+1)$ and 
$(Z,N)$. Each isotopic chain is connected to its neighbors by $\beta$-decay, and thus the $\beta$-decay lifetimes of 
nuclei along the $r$-process path set their relative abundances. Modern nuclear network calculations show that this 
equilibrium picture is an excellent approximation for early-time $r$-process evolution in many astrophysical 
scenarios. Eventually \nggn \ equilibrium fails, or in some scenarios is not established at all, 
and then neutron capture, photodissociation, and $\beta$-decay all compete to shape the final abundance pattern.

The roles of these individual pieces of data in $r$-process nuclear network simulations have been examined via 
sensitivity studies. In these studies, baseline astrophysical conditions are chosen, a single nuclear property is 
varied, and the simulation is rerun with the nuclear data change and compared to the baseline. Sensitivity studies 
highlight the pieces of data with the most leverage on the final abundance pattern and elucidate the mechanisms of 
influence. They have so far been performed for neutron capture rates 
\cite{Beun2009,Surman2009,Mumpower2012c,INPC2013}, $\beta$-decay rates \cite{INPC2013,AIP-Beta}, and 
photodissociation rates via their dependence on nuclear masses \cite{Brett2012,ICFN5,AIP-BE}, and are in progress for 
$\beta$-delayed neutron emission probabilities \cite{ARIS}. These studies are distinct from and complementary to 
studies of the influence of groups of nuclear properties, e.g. \cite{Suzuki2012,Nishimura2012}, global theoretical 
model predictions, e.g. \cite{Arcones2011}, or global variations in a Monte Carlo approach 
\cite{Mumpower2014_CGS15,Mumpower2014F} that can quantify correlations between nuclear physics inputs 
\cite{Bertolli2013}.

The sensitivity studies described above looked for the individual nuclear properties with the greatest impact on the $r$ 
process by considering variations of one piece of data at a time. However, we know that modifying a single nuclear mass 
alters all of the nuclear properties that depend on that mass. Here we perform nuclear mass $r$-process sensitivity 
studies in which variations in individual nuclear masses are consistently propagated to all affected nuclear properties, 
including neutron capture rates, photodissociation rates, and $\beta$-decay properties. This approach was developed in 
Ref.~\cite{Mumpower2015a} for spherical nuclei and is extended for the first time in this work to the entire chart of the 
nuclides between $Z=30$ and $Z=80$. These new studies capture the full impact of the uncertainties in individual masses 
on the $r$ process and point to the most important masses to measure in present and future experimental campaigns. 
Additionally they place direct constraints on the precision needed for measurements of nuclear properties and theoretical 
models, in order to improve $r$-process predictions and, eventually, distinguish between possible astrophysical sites.

\section{Baseline $r$-process simulations}
\label{sec:baseline}

We begin our sensitivity studies with a choice of baseline astrophysical trajectories. We investigate a range of 
astrophysical conditions including a low entropy hot wind, high entropy hot wind, a cold wind and a neutron star merger. 
In a hot wind an equilibrium is established between neutron captures and their inverse reaction, photodissociation, and 
the $\beta$-decays that move the nuclear flow to higher atomic number, $Z$, control the timescale for heavy-element 
production. Nuclear masses are influential during equilibrium as they directly set the $r$-process path for a given 
temperature and density, as shown in Eqn.~\ref{eq:saha}. To explore nuclear mass uncertainties in a low entropy hot wind 
we use a parameterized wind model from \cite{Meyer2002} which has an entropy of $30$ $k_B$, an electron fraction of 
$Y_e=0.20$ and a timescale of $70$ ms. This trajectory yields the production of heavy elements out to the third peak but 
is not neutron rich enough to lead to fission recycling. The high entropy hot trajectory uses the same parameterization 
with entropy of $100$ $k_B$, an electron fraction of $Y_e=0.25$ and a timescale of $80$ ms. In a cold wind the 
\nggn \ equilibrium is short-lived as the temperature drops quickly and photodissociation becomes 
negligible. The $r$-process path moves far from stability where a new quasi-equilibrium can be established between 
neutron captures and $\beta$-decays. To study nuclear masses under these conditions we employ a neutrino-driven wind 
simulation of Ref. \cite{Arcones2007} with artificially reduced electron fraction of $Y_e=0.31$ to produce a main $r$ 
process. We also consider a neutron star merger environment using a trajectory from Bauswain and Janka, for which the 
mildly-heated ejecta is sufficiently neutron-rich to undergo fission recycling \cite{Goriely2011}.

In all cases, once the supply of free neutrons is consumed the $r$-process path begins to move back to stability. The 
criterion of neutron exhaustion signals the start of the freeze-out phase of the $r$ process in which key abundance 
features are formed, such as the rare earth peak \cite{Surman1997,Mumpower2012b}. Additional neutrons during this 
time come from photodissociation, neutrons emitted promptly after $\beta$-decay, or fission \cite{Mumpower2012a}. 
During this freeze-out phase the reaction flows are sensitive to individual masses of nuclei between the $r$-process 
path and stability.

Our studies employ a dedicated $r$-process reaction network code \cite{Surman1997,Surman2001} which has been updated in 
recent studies \cite{Surman2008,Mumpower2012a} and includes a schematic treatment of fission \cite{Beun2008}. For the 
baseline nuclear masses we use the 2012 version of the Finite-Range Droplet Model (FRDM2012) \cite{FRDM2012-temp}. 
This model has an root-mean-square (rms) error of $0.57$ MeV compared to known masses from the Atomic Mass Evaluation (AME2012) \cite{AME2012}. 
When available, we use measured masses from the AME and keep these masses unchanged during our calculations. The use of experimental (AME2012) and theoretical (FRDM2012) mass values requires some care to avoid large discontinuities. Therefore when calculating quantities that depend on two masses of 
different origin (theory and experiment), we make sure to consistently calculate that quantity within a given dataset. 
For example, if a measured mass is available for $(Z+1,N-1)$ but not $(Z,N)$, the $\beta$-decay $Q$-value for $(Z,N)$ is 
calculated from the FRDM2012 masses only.

The baseline neutron capture rates are calculated with the publicly available statistical model code TALYS \cite{TALYS} using FRDM2012 and the latest compilation of measured masses from the AME2012 as described above. 
We use the default settings for level density, $\gamma$-strength function and particle optical model. The default level density model used in TALYS combines a 
constant nuclear temperature at low energies and matches it with a back-shifted Fermi gas model using systematics 
\cite{CT}. 
Masses can enter into this level density model via the definition of the constant temperature, which, as used in the TALYS code, is proportional to one over the square root of the shell correction term: $dW(Z,N) = M(Z,N)-M_{LDM}(Z,N)$ where $M(Z,N)$ is the mass of the nucleus and $M_{LDM}(Z,N)$ is the predicted mass to a spherical liquid-drop. 
Additionally, the level density parameter, $a$, is proportional to $dW$ and hence the nuclear masses. 
The default $\gamma$-strength function used in TALYS is the formulation set out in Kopecky-Uhl (KU) \cite{KU}. 
Nuclear masses enter into this parameterization of the giant dipole resonance via nuclear temperature term which prevents the $\gamma$ strength from going to zero as $\gamma$ energy decreases. 
The nuclear temperature itself is proportional to the square root of the neutron separation energy, $S_n$, as well as $a$, evaluated at $S_n$.

Photodissociation rates are calculated from neutron capture rates by detailed balance: 
\begin{equation}\label{eqn:photo}
\lambda_\gamma(Z,N) \propto T^{3/2} \exp\left[-{\frac{S_n(Z,N)}{kT}}\right] \langle \sigma v \rangle_{(Z,N-1)}
\end{equation}
where $S_n(Z,N)= M(Z,N-1)-M(Z,N)+M_n$ is the one neutron separation energy, $M(Z,N)$ and $M(Z,N-1)$ are masses of the 
nuclides, $M_n$ is the mass of the neutron, $T$ is the temperature, $\langle \sigma v \rangle_{(Z,N-1)}$ is the 
neutron capture rate of the neighboring nucleus and $k$ is Boltzmann's constant. The baseline photodissociation rates 
use FRDM2012 \& AME2012 masses and the neutron capture rates as calculated above. 

For the baseline $\beta$-decay rates we use experimental values \cite{NNDC} where available and theoretical estimates 
everywhere else. Theoretical rates are calculated as in \cite{Moller2003}:
\begin{eqnarray}\label{eqn:weak}
\lambda_\beta\equiv\frac{\ln(2)}{t_{1/2}}=\sum_{i} f^I_{\omega_i} C^{I}(\omega_i)
\end{eqnarray}
where $\lambda_\beta$ is the $\beta$-decay rate, $t_{1/2}$ is the half-life, $i$ denotes the $i$th excited state of 
the daughter nucleus with energy $E_i$, $\omega_i=(Q_\beta-E_i)/m_e$ is the $\beta$-decay energy to this state in 
units of electron mass, $I$ is the type of the decay, either Gamow-Teller (GT) or First-Forbidden (FF), $f$ is the 
phase space factor and $C$ is the $\beta$-strength function. We use the $\beta$-decay strength data from 
\cite{Moller2003} and calculate the phase space piece using FRDM2012 \& AME2012 masses as described above. Neutron 
emission probabilities are calculated by combining the Quasiparticle Random-Phase Approximation (QRPA) model of Ref. 
\cite{Moller2003} with a Hauser-Feshbach (HF) model from Ref. \cite{Kawano2008} which has been extended to the 
neutron dripline in this work. This combined QRPA-HF approach produces larger predictions of average neutron emission 
towards the dripline compared to older (QRPA) methods, as it follows the statistical decay until the initial 
available excitation energy is exhausted.

\section{Mass variation propagation}
\label{sec:dM}

For the sensitivity studies of this work we consider mass variations of $\pm0.5$ MeV, approximately equal to the rms 
error of FRDM2012. Since measured masses tend to have experimental uncertainties much smaller than this, we restrict 
our individual mass variations to nuclei with extrapolated or unknown masses.

\begin{figure}
 \begin{center}
  \centerline{\includegraphics[width=85mm]{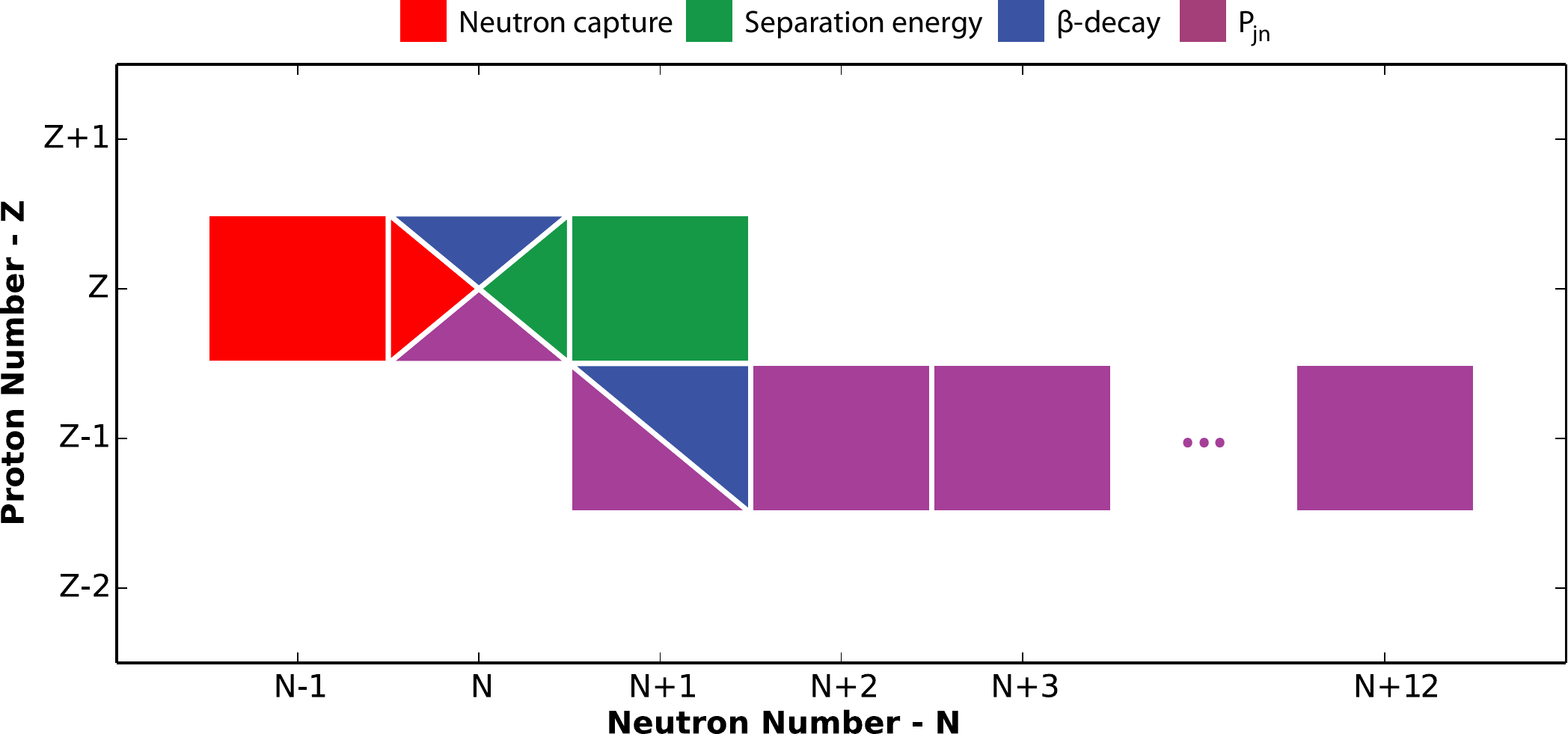}}
  \caption{\label{fig:mods} Shows the quantities of neighboring nuclei of importance to the $r$-process that may be altered by a change in mass of nucleus with $Z$ protons and $N$ neutrons.}
 \end{center}
\end{figure}

When an uncertain nuclear mass is varied we \textbf{recalculate} all the relevant nuclear properties of neighboring 
nuclei that depend on the changed mass. Specifically, if the mass of a nucleus ($Z$,$N$) with $Z$ protons and $N$ 
neutrons is varied then it can lead to changes in the neutron capture rates of ($Z$,$N$) and ($Z$,$N-1$), the 
separation energies of ($Z$,$N$) and ($Z$,$N+1$), the $\beta$-decay rates of ($Z$,$N$) and ($Z-1$,$N+1$), and 
$\beta$-delayed neutron emission probabilities of ($Z$,$N$), ($Z-1$,$N+1$), ($Z-1$,$N+2$), up to ($Z-1$,$N+12$) as 
shown in Fig.~\ref{fig:mods}.

To propagate the changes to the neighboring neutron capture rates we invoke the `massnucleus' command in TALYS and 
continue to use the default settings for level density, $\gamma$-strength function and particle optical model with 
the varied mass. A $\pm0.5$ MeV mass variation results in a change in neutron capture rates of a factor of 
approximately two to five. Note that this corresponds to the uncertainty in the neutron capture rates due only to 
uncertain nuclear masses; larger variations in the rates come from choice of statistical model components and the 
treatment of direct capture \cite{Beard2014}. The impact on the $r$ process of these larger neutron capture rate 
uncertainties are addressed in earlier studies \cite{Surman2001,Surman2009,Mumpower2012c}.

For the $\beta$-decay rates, all measured half-lives remain unchanged during our mass variations and we adjust only 
the affected theoretical rates. For the theoretical values, the majority of the dependence on nuclear mass is 
contained in the phase space piece, which goes as $f(\omega)\sim \omega^5$ for allowed decays. This is a 
significantly stronger dependence on masses than appears in the nuclear matrix elements calculated in the QRPA 
\cite{Mumpower2015a}. Therefore when updating the theoretical rates with the mass variations we recalculate only the 
phase space factors and leave the $\beta$-strength functions unchanged. This greatly reduces the numerical cost of 
the calculation while still capturing most of the dependence of the half-lives on the masses. We find mass variations 
of $\pm0.5$ MeV lead to changes in the half-lives by a factor of roughly two to four.

We recalculate the affected $\beta$-delayed neutron branching ratios in our QRPA-HF approach, assuming the structure 
of the ground state does not change with the mass variation. Modifications come from a change in the $\beta$-decay 
$Q$-value of the parent nucleus or from a change to the neutron separation energies of the daughter nuclei. A general 
dependence of neutron emission probabilities on changes in mass using QRPA-HF is thus entangled with the inclusion of 
$\gamma$-ray competition at each neutron emission stage.

\section{Mass sensitivity study}
\label{sec:sens}

With the required baseline nuclear data generated as described in Sec.~\ref{sec:baseline} and the computational tools in place 
to propagate the mass variations to the affected nuclear properties as described in Sec.~\ref{sec:dM}, we begin our 
sensitivity studies with the baseline astrophysical trajectories identified in Sec.~\ref{sec:baseline}. For each 
nuclear species $(Z,N)$ we repeat the baseline $r$-process simulation twice: once where the mass $M(Z,N)$ is 
increased by $0.5$ MeV and once where the mass is decreased by $0.5$ MeV. We then compute the metric
\begin{equation}
\label{eqn:F} 
F=100\sum_{A}|X(A)-X_{b}(A)| 
\end{equation} 
where $X_{b}(A)$ is the final isobaric mass fraction in the baseline simulation, $X(A)$ is the final isobaric mass 
fraction of the simulation when all nuclear inputs have been modified based off the change in a single mass, and the 
summation runs over the entire baseline pattern \cite{Mumpower2015a}. As defined here, the $F$ metric quantifies the 
impact of the mass on the \emph{global} $r$-process pattern. This procedure is then repeated for every nuclear 
species from the limits of the AME2012 experimental values to the FRDM2012 neutron drip line.

\begin{figure*}
 \begin{center}
  \centerline{\includegraphics[width=160mm]{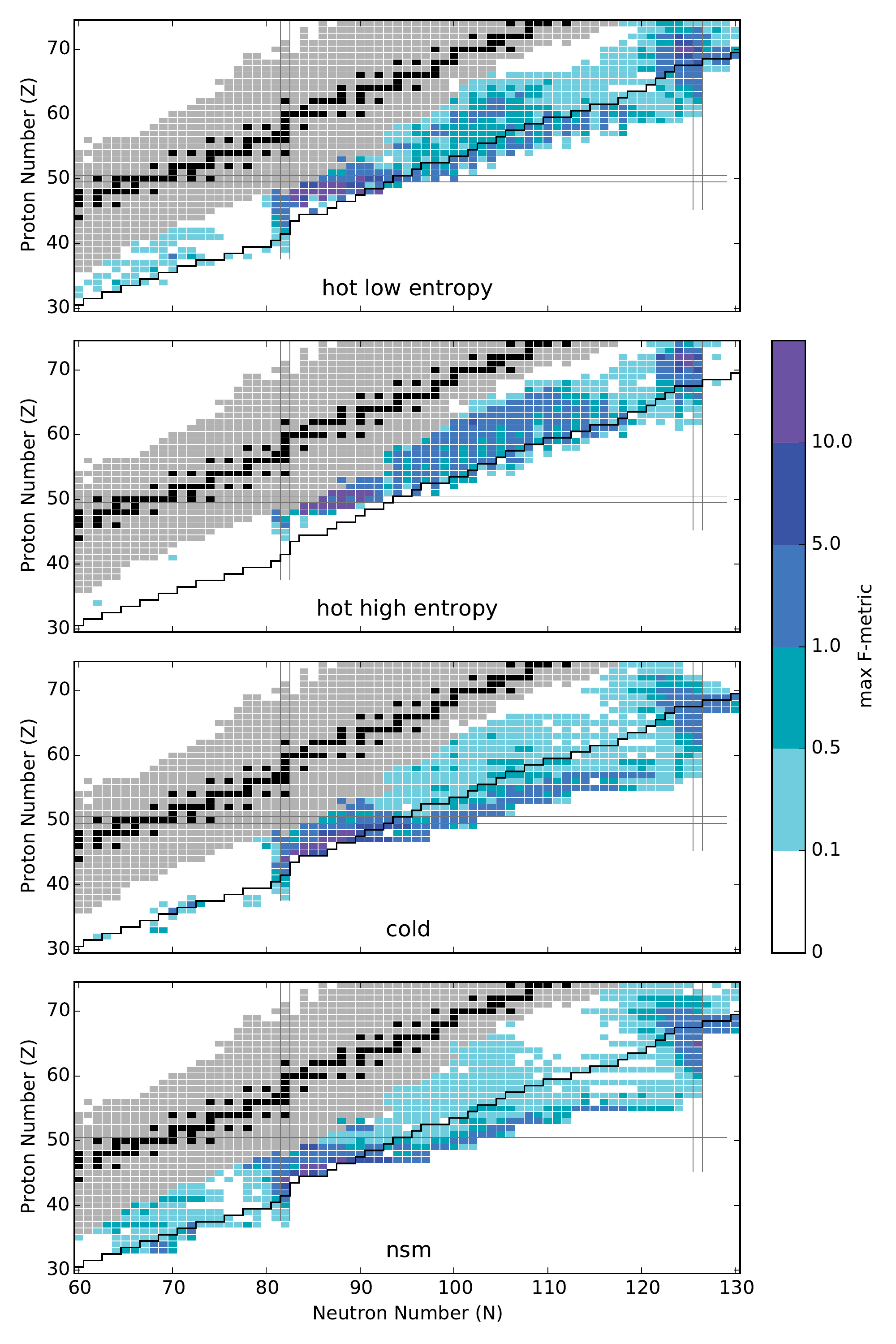}}
  \caption{\label{fig:fgrid} Nuclei that significantly impact final $r$-process abundances for four astrophysical 
conditions. The intensity of the color denotes the maximum $F$ value, Eqn.~\ref{eqn:F}, resulting from individual $\pm0.5$ MeV 
mass variations. Light gray denotes the extent of measured masses from the 2012 AME and stable nuclei are colored black. 
For reference, estimated accessibility limits are shown for the upcoming Facility for Rare Isotope Beams (FRIB) (black line - intensity of $10^{-4}$ particles per second \cite{FRIB}).}
 \end{center}
\end{figure*}

Fig.~\ref{fig:fgrid} shows the results of the full-chart mass sensitivity studies run for our four choices of astrophysical conditions. 
The shading of each rectangle represents the largest of the two $F$ values that we obtained when we calculated the $r$-process abundances for the reference mass (FRDM2012) (1) plus $0.5$ MeV and (2) minus $0.5$ MeV. 
Roughly 1100 nuclei are included in each study with $30 \leq Z \leq 80$ and $60 \leq N \leq 130$. 
As a general result, we find influential nuclear masses lie along the equilibrium $r$-process path as well as along the decay pathways back to stability, most of which are within the predicted experimental reach of the upcoming Facility for Rare Isotope Beams (FRIB) \cite{Thoennessen2014}. 
This is particularly evident around the closed neutron shells where we find the largest global impact of nuclear masses. 

Qualitatively, the overall distribution of influential nuclear masses in the four cases reinforces the conclusions 
from our previous work \cite{Brett2012,AIP-BE,Mumpower2015a}. In particular the pattern of most impactful nuclei for 
the hot wind $r$ process shown in the top panel of Fig.~\ref{fig:fgrid} is directly comparable to that from 
\cite{Brett2012,AIP-BE}. Our new studies show several key improvements, which come from addressing the limitations of 
these early studies. Refs.~\cite{Brett2012,AIP-BE} considered the propagation of mass variations to photodissociation 
rates only, which underestimates the resulting sensitivity measures, especially in cold wind or merger $r$-process 
scenarios where photodissociation is suppressed. In these cases, the influence of masses on the $r$ process occurs 
via the decay properties and neutron capture rates---an effect which was captured for the first time in 
Ref.~\cite{Mumpower2015a} but only for closed shell nuclei. Thus our new studies are the first reliable full-chart 
estimates of the impact of nuclear masses in cold and merger $r$-process scenarios. In addition, the hot wind studies 
show higher mass sensitivities in the rare earth region compared to the earlier studies, due to the interplay of 
$\beta$-decay, neutron capture, and photodissociation that forms the rare earth peak \cite{Surman1997,Mumpower2012a}. 
The exact mechanisms by which each piece of nuclear data influences the final abundance pattern in different types of 
$r$ processes are explored in 
Refs.~\cite{Beun2009,Surman2009,Mumpower2012c,AIP-Beta,Brett2012,AIP-BE,ARIS,Mumpower2015a}.

Any nucleus with impact parameter above $F\sim10$ (purple shading) implies a significant global change to the 
abundances. Only Palladium ($_{46}$Pd), Cadmium ($_{48}$Cd), Indium ($_{49}$In), and Tin ($_{50}$Sn) isotopes show 
sensitivities above $F=20$ in our studies. Table 1 includes the $F$ measures of these most impactful nuclei.
\begin{longtable}{@{\extracolsep{\fill}} ccccc }
 \caption{\label{tab:F} Important nuclei from Fig.~\ref{fig:fgrid} with $F_{max}\geq20$. }
 \\
 \multicolumn{1}{c}{$Z$} & \multicolumn{1}{c}{$N$} & \multicolumn{1}{c}{$A$} & \multicolumn{1}{c}{$F_{max}$} & \multicolumn{1}{c}{Trajectory} \\
 \hline\hline
 \endhead
 48 & 84 & 132 & 101.39 & high entropy hot \\ 
 50 & 86 & 136 & 83.68 & high entropy hot \\ 
 49 & 84 & 133 & 74.59 & high entropy hot \\ 
 49 & 86 & 135 & 73.82 & high entropy hot \\ 
 49 & 85 & 134 & 72.84 & high entropy hot \\ 
 49 & 87 & 136 & 71.04 & high entropy hot \\ 
 49 & 88 & 137 & 68.08 & high entropy hot \\ 
 49 & 89 & 138 & 66.43 & high entropy hot \\ 
 48 & 88 & 136 & 58.42 & low entropy hot \\ 
 46 & 86 & 132 & 55.56 & nsm \\ 
 48 & 86 & 134 & 52.38 & low entropy hot \\ 
 46 & 86 & 132 & 48.67 & cold \\ 
 46 & 84 & 130 & 43.73 & cold \\ 
 48 & 90 & 138 & 37.86 & low entropy hot \\ 
 46 & 85 & 131 & 34.81 & cold \\ 
 50 & 88 & 138 & 29.64 & high entropy hot \\ 
 46 & 84 & 130 & 24.19 & nsm \\ 
 48 & 88 & 136 & 24.10 & cold \\ 
 48 & 85 & 133 & 23.38 & low entropy hot \\ 
 48 & 87 & 135 & 21.12 & low entropy hot \\ 
 48 & 89 & 137 & 20.63 & cold \\ 
 \hline
\end{longtable} 

\section{Abundance pattern predictions}

\begin{figure}
 \begin{center}
  \centerline{\includegraphics[width=95mm]{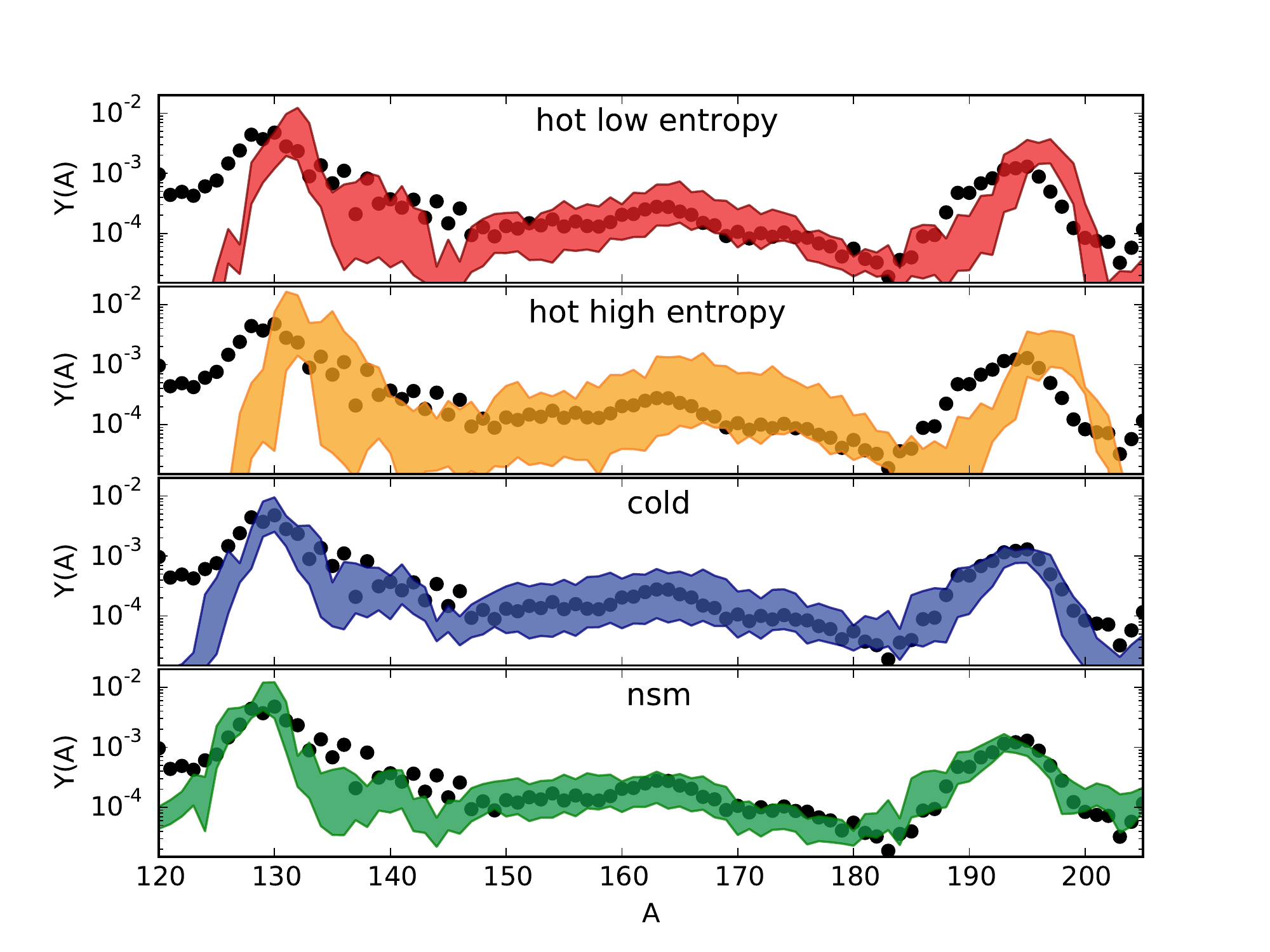}}
  \caption{\label{fig:ab} Variances of the ensembles of final abundance patterns (shaded bands) for the four sensitivity 
studies described in this work, compared to scaled solar $r$-process residuals (black circles) from 
\cite{Arlandini1999}.}
 \end{center}
\end{figure}

We now use the above sensitivity study results to examine the variances that can arise in the final abundance pattern 
due to uncertainties in nuclear masses. In each sensitivity study we first select the nuclei for which $F\geq0.1$ as 
shown in Fig.~\ref{fig:fgrid}. We create an ensemble of abundance patterns from the simulations that include the 
individual mass changes $\pm0.5$ MeV of these nuclei. We then compute the variance of this reduced ensemble of 
abundances for each value of $A$. The variance in the final abundance patterns for the low entropy hot wind, high 
entropy hot wind, cold wind and neutron star merger studies are shown in Fig.~\ref{fig:ab} compared to the solar 
isotopic $r$-process residuals.

The dependence of $r$-process predictions on the uncertainties in nuclear masses is shown by variance bands for four different astrophysical conditions in Fig.~\ref{fig:ab}. 
These four trajectories have distinct $r$-process paths and dynamics during freeze-out which means the variance in abundances in each environment comes from a different aspect of the dependence on nuclear masses as mentioned above. 
This result is only obtainable by using our approach of consistently propagating uncertainties from nuclear masses to all of the relevant nuclear quantities for the $r$ process. 

One may be tempted to rule out the hot conditions as the variance bands are clearly offset for the third ($A=195$) peak. 
However, we note that our approach here actually underestimates these bands \textit{in all scenarios} --- larger uncertainties from reaction rate calculations are not included, which are particularly important for the hot scenarios, and mass changes in these studies are done on an individual basis. 
Methods that rely on a global Monte Carlo approach have the ability to resolve these drawbacks. 
Preliminary work \cite{Mumpower2014_CGS15} in this direction suggests details of the abundance pattern can be clearly resolved if mass uncertainties are reduced to less than $0.1$ MeV. 
The next generation of nuclear mass measurement campaigns will be crucial in the progress toward this ambitious goal. 

With this caution in mind, Fig.~\ref{fig:ab} confirms our previous results that mass model uncertainties are currently too large for precision abundance pattern predictions capable of differentiating between $r$-process conditions \cite{Mumpower2015a}. 
For example, the formation of the $A\sim 160$ rare earth peak can in principle be used to constrain the $r$-process site \cite{Mumpower2012b}, however the variance bands in Fig.~\ref{fig:ab} are larger than the peak itself. 
This indicates that the features of the mass surface in this region responsible for rare earth peak formation are likely on the order of the rms value of FRDM2012 or smaller. 

We also note that the new FRDM2012 masses show marked improvement over the FRDM1995 masses in matching features from the solar pattern, as is clear from a comparison of the final abundances of Fig. \ref{fig:ab} to those in Ref.~\cite{Mumpower2015a}. 
The most notable improvement of the mass model comes from an enhanced description of nuclei in the transition region between the $N\sim82$ region and the rare earth region.

\section{Conclusion}

In summary, we have shown for the first time how uncertainties in \textit{individual} nuclear masses propagate to 
influence and shape the $r$-process abundance distribution across the chart of nuclides. We consider variations of 
individual nuclear masses and recalculate consistently all relevant $Q$-values, neutron capture rates, 
photodissociation rates, $\beta$-decay rates and $\beta$-delayed neutron emission probabilities, as shown in Fig. 
\ref{fig:mods}. We find mass uncertainties of $\pm0.5$ MeV have a significant impact on $r$-process abundance 
predictions as summarized in Fig.~\ref{fig:fgrid}. In terms of our metric, a value of $F\sim20$ represents a large 
local change or a global shift in $r$-process abundances.

We explore changes to masses of $\pm0.5$ MeV from FRDM2012 in four astrophysical trajectories: a low entropy hot 
wind, a high entropy hot wind, a cold wind, and a neutron star merger. Shifts in the equilibrium path play the 
dominant role in a hot $r$ process, and our results here mirror earlier studies \cite{Brett2012,AIP-BE} where mass 
variations were propagated only to the photodissociation rates. Changes to weak decay properties and neutron capture 
rates are essential to include particularly in the cold wind and merger cases where photodissociation channel is 
suppressed. We find a similar dependence on nuclear masses for all astrophysical conditions studied, as shown in Fig. 
\ref{fig:ab}, due to the propagation of mass uncertainties to all relevant quantities. The nuclei with the most 
impactful masses lie along the equilibrium $r$-process path, as expected, and also along the decay paths to 
stability. This result strongly reinforces our conclusions from previous studies that understanding freeze-out is 
critical for predicting $r$-process abundances under any astrophysical conditions.

Many of the influential nuclear masses identified in our studies will be accessible to future radioactive beam facilities. 
These measurements have the potential to dramatically improve the precision of $r$-process simulations. 
Still, some masses will remain beyond experimental reach. 
It is therefore also of key importance to improve global mass models. 
This work provides additional motivation for strengthening efforts in improving the global description of nuclear masses as it clearly shows that current mass model uncertainties are still too large for the finer details of the isotopic $r$-process abundance pattern to be resolved. 
The question then arises: how well do we need to know nuclear masses in order to predict a final composition of the $r$ process? 
Studies which address this question strongly point to the resolution of abundance features if global model uncertainties are reduced to less than $0.1$ MeV \cite{Mumpower2015a,Mumpower2014_CGS15}. 
Thus, a concerted effort is required by the community to analyze model inputs and physical assumptions in order to achieve reliable calculations as close to this threshold as possible \cite{Bohigas2002,Barea2005,Olofsson2006}. 

\section{Acknowledgements}

We thank Gail McLaughlin for helpful discussions. This work was supported in part by the National 
Science Foundation through the Joint Institute for Nuclear Astrophysics grant numbers PHY0822648 and PHY1419765, and the 
Department of Energy under contract DE-SC0013039 (RS).

\bibliographystyle{apsrev4-1}
\bibliography{refs}

\end{document}